\documentstyle [twoside,12pt]{article}
\newcommand{\nc}{\newcommand}
\nc{\la}{\lambda} \nc{\al}{\alpha}
\nc{\th}{\theta}  \nc{\be}{\beta}
\nc{\ga}{\gamma}  \nc{\Ga}{\Gamma}
\nc{\de}{\delta}  \nc{\De}{\Delta}
\nc{\si}{\sigma}  \nc{\ka}{\kappa}
\nc{\om}{\omega}  \nc{\Om}{\Omega}
\nc{\nf}{\infty}
\nc{\ra}{\rightarrow}
\nc{\beq}{\begin{equation}}
\nc{\eeq}{\end{equation}}
\nc{\beqa}{\begin{eqnarray}}  \nc{\dst}{\displaystyle}
\nc{\eeqa}{\end{eqnarray}} \nc{\nnb}{\nonumber}
\title{ \bf Anomalies in \\ N=2 supersymmetric non-linear $\si$ models \\ on
compact K\"ahler Ricci-flat target spaces }
\author{Guy Bonneau\thanks
{\noindent Laboratoire de Physique Th\'eorique et des Hautes Energies,
 Unit\'e associ\'ee au CNRS URA 280,~Universit\'e Paris 7,
 2 Place Jussieu, 75251 Paris Cedex 05.}}
\begin{document}
\maketitle
\begin{abstract}
\noindent We analyse with the algebraic, regularisation independent,
cohomological B.R.S. methods, the renormalisability of torsionless N=2
supersymmetric non-linear $\si$ models built on K\"ahler spaces. Surprisingly
enough with respect to the common wisdom, we obtain an anomaly candidate, at
least in the compact Ricci-flat case. In the compact homogeneous K\"ahler case,
as expected, the anomaly candidate disappears.
\end{abstract}

\vfill
{\bf PAR/LPTHE/94-09}\hfill  Mars 1994

\section{Introduction}
Supersymmetric non-linear $\si$  models in two space time dimensions have been
considered for many years to describe the vacuum state of superstrings
\cite{1}\cite{1a}. In particular Calabi-Yau spaces, {\it i.e.} 6 dimensional
compact K\"ahler Ricci-flat manifolds \cite {2}, appear as good candidates in
the compactification of the 10 dimensional superstring to 4 dimensional flat
Minkowski space ; the conformal invariance of the 2.d, N = 2 supersymmetric
non-linear $\si$  model (the fields of which are coordinates on this compact
manifold) is expected to hold to all orders of perturbation theory \cite {3}.

However explicit calculations to 4 or 5 loops \cite {4} and, afterwards,
general arguments \cite{5} show that the $\be$ functions may not vanish.
But, as argued in my recent review \cite{6}, at least two problems obscure
these analyses : first, the fact that the quantum theory is not sufficiently
constrained by the K\"ahler Ricci-flatness requirement ; second, the use of
``dimensional reduction" \cite{62} or of harmonic superspace formalism
\cite{61} \footnote{\ The regularisation through dimensional reduction suffers
from algebric inconsistencies and the quantization in harmonic superspace does
not rely on firm basis, due to the presence of non-local singularities ( in the
harmonic superspace) \cite{63}.} in actual explicit calculations and general
arguments. Then, we prefer to analyse these models using the B.R.S., algebraic,
regularisation free cohomological methods.

So we adress ourselves to the question of the all-order renormalisability of
supersymmetric (N = 2, 4) non linear $\si$   models in two space time
dimensions, in the same spirit of what we did with the Genova group \cite{7}
for the bosonic case. Due to the non-linearity of the second supersymmetry
transformation in a general field parametrisation (coordinate system on the
manifold), we shall use a gradation in the number of fields and their
derivatives. As in \cite {7}, the cohomology of the lowest order B.R.S.
operator will give the essential information. Leaving to other publications the
detailed analysis of the N=2 supersymmetry and the N=4 case \cite {13}, the
present letter gives our main results for \underline{torsionless} compact
K\"ahler Ricci-flat manifolds ({\it i.e.} special N=2 supersymmetric models)
and, surprisingly enough with respect to the common wisdom \footnote{\ Notice
also that recent works of Brandt \cite{8} and Dixon \cite{9} show the existence
of new non-trivial cohomologies in supersymmetric theories.}, shows that, at
least for that case, there exists a possible anomaly for \underline{global}
supersymmetry in 2 space-time dimensions.

\section {The classical theory and the Slavnov operator}

As in this letter we shall be concerned in N = 2 supersymmetric non-linear
$\si$  models in 2.d, we may use N = 1 superfields \footnote{\ The quantization
with N = 1 superfields was put on firm basis by Piguet and Rouet \cite{10} who
proved in particular the Quantum Action Principle in that context. Moreover, in
\cite{13}a) we show the renormalisability of N=1 supersymmetric non-linear
$\si$ models using component fields : this justifies the use of N=1 superfields
for the present analysis of extended supersymmetry. \newline Notice also that
we use light-cone coordinates.  } $\phi^i (x^+,x^-,\th^+,\th^-)$  and
consequently, in the  \underline{absence of torsion}, the most general N = 1
invariant action is :
\beq\label{a1}
S^{inv.} = \int d^2x d^2\th g_{ij}[\phi]D_+\phi^iD_-\phi^j
\eeq
where the supersymmetric covariant derivatives $$ D_{\pm} =
\frac{\partial}{\partial \th^{\pm}} + i\th^{\pm} \frac{\partial}{\partial
x^{\pm}}  $$ satisfy $$ \{ D_+ , D_- \} = 0 \ \ \ ,\ \ D^2_{\pm} =
i\frac{\partial}{\partial x^{\pm}} \equiv i\partial _{\pm} \ \ .$$

\noindent The tensor $ g_{ij}[\phi]$ is interpreted as a metric tensor on a
Riemannian manifold $\cal M$ \footnote {\ Here, contrarily to our previous work
where the manifold $\cal M$ was supposed to be an homogeneous space \cite {7},
we consider renormalisability ``{\it \`a la Friedan} " \cite {100}, {\it i.e.}
in the space of metrics, and analyse only the possibility of maintaining to all
orders the N=2 supersymmetry. As explained in \cite {6}, in order to define
unambiguously the classical action, one should add extra properties. }. As is
by now well known \cite {11}, N = 2 supersymmetry needs $\cal M$ to be a 2n
dimensional K\"ahler manifold, {\it i.e.} there should exist a covariantly
constant complex structure $J^i_j [\phi]$ :
$$ J^i_jJ^j_k = -\de^i_k \ \ \ ; \ \nabla_kJ^i_j = 0 \ \ ;\ \  i,j,k =
1,2,....2n $$
and the metric has to be hermitian with respect to the complex structure :
$$g_{kl}J^l_n \ +\ J^l_k g_{ln} = 0 \ \ . $$
The second supersymmetry transformation writes :
\beq\label{a2}
\de\phi^i \ =\ J^i_j[\phi](\epsilon^+ D_+\phi^j + \epsilon^- D_-\phi^j)\ .
\eeq

In the B.R.S. approach, the supersymmetry parameters $\epsilon^{\pm}$ are
promoted to constant, commuting Faddeev-Popov parameters $d^{\pm}$   and an
anticommuting classical source $\eta_i$  for the non-linear field
transformation (\ref{a2}) is introduced in the classical action \footnote {\ In
the absence of torsion, there is a parity invariance $$ + \ra -, d^2x \ra d^2x,
d^2\th \ra -d^2\th, \phi^i \ra \phi^i, \eta_i \ra -\eta_i \ .$$ Moreover, the
canonical dimensions of $[ d^2x d^2\th],\ [\phi^i],\ [d^{\pm}],\ [D_{\pm}],\
[\eta_i]$ are -1, 0, -1/2, +1/2, +1 respectively and the Faddeev-Popov
assignments + 1 for $d^{\pm}$, -1 for $\eta_i$, 0 for the other quantities.}:
\beq\label{a3}
\Ga^{class.} = \int d^2x d^2\th \left\{ g_{ij}[\phi]D_+\phi^iD_-\phi^j + \eta_i
J^i_j[\phi](d^+ D_+\phi^j + d^- D_-\phi^j) \right\}\ .
\eeq
For simplicity, no mass term is added here as we are only interested in U.V.
properties.
The non linear Slavnov operator is defined by
$$S\Ga \equiv \int d^2x d^2\th \frac {\de\Ga}{\de\eta_i(x,\th)}\frac
{\de\Ga}{\de\phi^i(x,\th)} $$
and we find $$S\Ga^{class.} = (d^+)^2\int d^2x d^2\th \eta_k
i\partial_{++}\phi^k + (d^-)^2\int d^2x d^2\th \eta_k i\partial_{--}\phi^k $$
in accordance with the supersymmetry algebra.

As is by now well known (for example see \cite{6} or \cite{7}), in the absence
of a consistent regularisation that respects all the symmetries of the theory,
the quantum analysis directly depends on the cohomology of the nihilpotent
linearized Slavnov operator :
\beqa\label{a4}
S_L &=& \int d^2x d^2\th \left[\frac {\de\Ga^{class.}}{\de\eta_i(x,\th)}\frac
{\de}{\de\phi^i(x,\th)} + \frac {\de\Ga^{class.}}{\de\phi^i(x,\th)}\frac
{\de}{\de\eta_i(x,\th)} \right] \nnb\\
S_L^2 &=& 0
\eeqa
in the Faddeev-Popov charge +1 sector [absence of anomalies for the N = 2
supersymmetry] and 0 sector [number of physical parameters and stability of the
classical action through renormalization]. Notice that the Slavnov operator
(\ref{a4}) is unchanged under the following field and source reparametrisations
:
\beq\label{a40}
\phi^i \ra \phi^i + \la W^i[\phi] \ \ \ ,\ \ \eta_i \ra \eta_i - \la \eta_k
W^k_{,i}[\phi] \ ,
\eeq
where $W^i[\phi]$ is an arbitrary function of the fields $\phi(x,\th)$ and a
comma indicates a derivative with respect to the field $\phi^i$. Under this
change, the classical action (\ref{a3}) is modified
\beq\label{a41}
\Ga^{class.} \ra \Ga^{class.} + \la S_L \int d^2x d^2\th \eta_i W^i[\phi]\ ,
\eeq
but the Slavnov identity is left unchanged as
$$ S[\Ga^{class.} + \la S_L\De] \equiv S\Ga^{class.} + \la S_L[S_L \De] =
S\Ga^{class.} \ .$$

\section{B.R.S. cohomology of $S_L$ }

Due to the highly non-linear character of $S_L$ (equ. (\ref{a4})), it is
convenient to use a ``filtration" (\cite{14},\cite{7}) with respect to the
number of fields $\phi^i(x,\th)$ and their derivatives. As it does not change
this number, the nihilpotent lowest order part of $S_L$, $S_L^0$ will play a
special role : :
\beqa\label{a5}
S_L &=& S_L^0 + S_L^1 + S_L^2 +... \equiv S_L^0 + S_L^r \ ,\ \ (S_L^0)^2 =
(S_L^r)^2 = S_L^0 S_L^r + S_L^r S_L^0 = 0 \nnb\\
S_L^0 &=&  \int d^2x d^2\th J^i_j(0) \left\{ (d^+ D_+\phi^j + d^-
D_-\phi^j)\frac{\de}{\de\phi^i} + (d^+ D_+\eta_i + d^-
D_-\eta_i)\frac{\de}{\de\eta_j}\right\}\ .
\eeqa

\noindent As explained in refs.\cite{7} and \cite{13}a), when $S_L^0$ has no
cohomology in the Faddeev-Popov positively charged sectors, the cohomology of
the complete $S_L$ operator in the Faddeev-Popov sectors of charge 0 and +1 is
isomorphic to a subspace\footnote {\  In particular, the cohomology of $S_L^0$
in the Faddeev-Popov -1 sector restricts the dimension of the cohomology of
$S_L$ in the 0 charge sector when compared to the one of  $S_L^0$.} of the one
of  $S_L^0$ in the same sectors.

We now analyse the cohomology of $S_L^0$.

\subsection{$S_L^0$ cohomology}
Due to dimensions and Faddeev-Popov charge assignments, dimension zero
integrated local polynomials in the Faddeev-Popov parameters, fields, sources
and their derivatives have at least a Faddeev-Popov charge -1 :
\beq\label{a7}
\De_{[-1]} = \int d^2x d^2\th \eta_iV^i[\phi]\ .
\eeq
Then there is no Faddeev-Popov charge -1 coboundaries, so the cohomology of
$S_L^0$ in that sector is given by the cocycle condition :
\beq\label{a8}
S_L^0 \De_{[-1]} = 0 \ \ \Leftrightarrow \ \ J^i_j(0)V^k_{,i} =
J^k_i(0)V^i_{,j}
\eeq
This condition, when expressed in a coordinate system adapted to the complex
structure  $J^i_j[\phi]$ ($J^{\al}_{\be} = i \de^{\al}_{\be} ,
J^{\bar{\al}}_{\bar{\be}} = -i \de^{\al}_{\be}, J^{\bar{\al}}_{\be} =
J^{\al}_{\bar{\be}} = 0 $), means that $V^i[\phi]$ is a contravariant analytic
vector : $V^{\al} = V^{\al}[\phi^{\de}] ,\ V^{\bar{\al}} =
V^{\bar{\al}}[\bar{\phi}^{\bar{\de}}]$.
$$ $$

Let us now turn to the Faddeev-Popov neutral charge sector :
\beq\label{a81}
\De_{[0]} = \int d^2x d^2\th \left\{ t_{ij}[\phi]D_+\phi^iD_-\phi^j + \eta_i
U^i_j[\phi](d^+ D_+\phi^j + d^- D_-\phi^j) \right\}
\eeq
where $t_{ij}$ is symmetric, due to parity invariance (footnote 5).
Coboundaries being given by $S_L^0 \De_{[-1]}$[arbitrary $V^i(\phi)$], the
analysis of the cocycle condition $S_L^0 \De_{[0]} = 0 $ gives
\beq\label{a811}
\De_{[0]} =  \De_{[0]}^{an.}[t_{ij}(\phi)] + S_L^0 \De_{[-1]}[V^i(\phi)]
\eeq
where the tensor  $t_{ij}$ which occurs in the anomalous part
\beq\label{a82}
\De_{[0]}^{an.}[t_{ij}] = \int d^2x d^2\th t_{ij}[\phi]D_+\phi^iD_-\phi^j
\eeq
is constrained by :
\beqa\label{a9}
a) & \ \ J^i_j(0)t_{ik} + t_{ji}J^i_k(0) &= 0 ,\nnb \\
b) & \ \ J^i_j(0)[t_{kl,i}-t_{il,k}] - (j \leftrightarrow k ) &= 0.
\eeqa

\noindent The absence of source dependent non-trivial cohomology means that, up
to a field redefinition (compare to equations (\ref{a40},\ref{a41})), the
complex structure $J^i_j$ is left unchanged through radiative corrections.
Moreover, using the same adapted coordinate system as above, condition
(\ref{a9}a)   means that the metric $g_{ij} + \hbar t_{ij}$ remains hermitian,
whereas (\ref{a9}b) expresses the covariant constancy of $J^i_j$ with respect
to the covariant derivative with a connexion corresponding to the metric
$g_{ij} + \hbar t_{ij}$. These are precisely the expected conditions.
$$ $$
Finally, let us consider the Faddeev-Popov charge +1 sector :
\beqa\label{a10}
\De_{[+1]} &=& \int d^2x d^2\th \{ t^{[ijk]}(d^+)^2(d^-)^2\eta_i\eta_j\eta_k
\nnb\\
&+& d^+d^-[\eta_i\eta_j t^{[ij]}_{1\; n}(d^+D_+\phi^n - d^-D_-\phi^n) + \eta_n
t^n_{2\; [ij]}D_+\phi^iD_-\phi^j ]\nnb\\
&+& d^+d^- \eta_i s^{(ij)}_1 (d^+D_+\eta_j - d^-D_-\eta_j) \nnb\\
&+& (d^+)^2(\eta_n t^n_{3\; [ij]}D_+\phi^iD_+\phi^j +
D_+\eta_iD_+\phi^jt^i_{4\;j})\nnb\\
&+& (d^-)^2(\eta_n t^n_{3\; [ij]}D_-\phi^iD_-\phi^j +
D_-\eta_iD_-\phi^jt^i_{4\;j})\nnb\\
&+& d^+(\tilde {t}_{[ij]n}D_+\phi^iD_+\phi^jD_-\phi^n + s_{2\;
(ij)}D_-D_+\phi^iD_+\phi^j)\nnb\\
&-& d^-(\tilde {t}_{[ij]n}D_-\phi^iD_-\phi^jD_+\phi^n + s_{2\;
(ij)}D_+D_-\phi^iD_-\phi^j)\}
\eeqa
where, due to the anticommuting properties of $\eta_i$ and $D_{\pm}\phi^i$ and
to the integration by parts freedom, the tensors $t^{[ijk]}$, $t^{[ij]}_{1\;
n}$, $t^n_{2\; [ij]}$, $t^n_{3\; [ij]}$, $\tilde {t}_{[ij]n}$ are antisymmetric
in i, j, k, and $s^{(ij)}_1$, $s_{2\; (ij)}$ symmetric in i, j.

\noindent Coboundaries being given by $S_L^0 \De_{[0]}$[arbitrary
$t_{ij}(\phi), U^i_j(\phi)]$, the analysis of the cocycle condition $S_L^0
\De_{[+1]} = 0$ leads to :
\beq\label{a91}
\De_{[+1]} =  \De_{[+1]}^{an.}[t^{[ijk]}(\phi)] + S_L^0 \De_{[0]}[t_{ij}(\phi),
U^i_j(\phi)]
\eeq
where the antisymmetric tensor $t^{[ijk]}(\phi)$ which occurs in the anomalous
part
\beq\label{a92}
\De_{[+1]}^{an.} = \int d^2x d^2\th
t^{[ijk]}(\phi)(d^+)^2(d^-)^2\eta_i\eta_j\eta_k
\eeq
is constrained by :
\beqa\label{a11}
a)& \ \ J^i_n(0)t^{[njk]} \ \ \ \ {\rm is\ \ i,\ j,\ k\ \ antisymmetric},\nnb\\
b)&  \ \ J^i_n(0)t^{[njk]}_{,m} = J^n_m(0)t^{[ijk]}_{,n}
\eeqa

\noindent Using the same adapted coordinate system as above, condition
(\ref{a11}a) means that the tensor  $t^{[ijk]}$ is a pure contravariant
antisymmetric tensor ({\it i.e.} $t^{[\al\be\ga]} ,
t^{[\bar{\al}\bar{\be}\bar{\ga}]} \neq 0$ , the other components vanish)
whereas (\ref{a11}b)    means that it is analytic ({\it i.e.} $t^{[\al\be\ga]}
= t^{[\al\be\ga]}(\phi^{\de})$,  $t^{[\bar{\al}\bar{\be}\bar{\ga}]} =
t^{[\bar{\al}\bar{\be}\bar{\ga}]}(\bar{\phi}^{\bar{\de}}))$. In particular, due
to the vanishing of $t^{[\al\be\bar{\ga}]}$,  such tensor cannot be a candidate
for a torsion tensor on a K\"ahler manifold \cite{151}.

Consider the covariant tensor
$$t_{[\al\be\ga]} =
g_{\al\bar{\al}}g_{\be\bar{\be}}
g_{\ga\bar{\ga}}t^{[\bar{\al}\bar{\be}\bar{\ga}]}\ .$$
It satisfies $\nabla _{\de}t_{[\al\be\ga]} = 0\ .$ Then the (3-0) form
$$\om ' = \frac{1}{3!}t_{[\al\be\ga]}d\phi^{\al}\wedge d\phi^{\be} \wedge
d\phi^{\ga}$$ which satisfies $d'\om' = 0$, may be shown to be harmonic if
$\cal{M}$ is a compact manifold
\footnote{\ In this K\"ahlerian case, one firstly obtains from $\nabla
_{\de}t_{[\al\be\ga]} = 0\ ,\ \ \ \triangle t_{[\al\be\ga]} = g^{\de\bar{\de}}
\nabla _{\de}\nabla _{\bar{\de}}t_{[\al\be\ga]} -[R^{\de}_{\al}t_{[\de\be\ga]}
+ \ {\rm perms.}\ ]\ $ ; on another hand, the Ricci identity gives
$g^{\de\bar{\de}} \nabla _{\de}\nabla _{\bar{\de}}t_{[\al\be\ga]} =
-[R^{\de}_{\al}t_{[\de\be\ga]} + \ {\rm perms.}\ ].$
So $ \triangle t_{[\al\be\ga]} = 2g^{\de\bar{\de}} \nabla _{\de}\nabla
_{\bar{\de}}t_{[\al\be\ga]}$. Now, the manifold being compact, one may compute
:
$$ (d\om ',d\om ') + (\de\om ',\de\om ') = (\om ',(d\de+\de d)\om ') = (\om
',\triangle\om ') = $$
$$= \int_{\cal{M}} d\si 2t^{[\al\be\ga]}g^{\de\bar{\de}} \nabla _{\de}\nabla
_{\bar{\de}}t_{[\al\be\ga]} = \int_{\cal{M}} d\si 2g^{\de\bar{\de}} \{\nabla
_{\de}\nabla _{\bar{\de}}(t^{[\al\be\ga]}t_{[\al\be\ga]}) - \nabla
_{\de}t^{[\al\be\ga]}\nabla _{\bar{\de}}t_{[\al\be\ga]}\} = 0 - 2(d\om ',d\om
')$$
$$\Rightarrow \ (\de\om ',\de\om ') + 3(d\om ',d\om ') = 0
\ \Rightarrow \ \de\om ' = d\om '= \triangle \om' = 0\ . \ \ \ Q.E.D.$$ }
(\cite{16},\cite{13}b)). It is known that the number of such forms is given by
the Hodge number $h^{3,0}$ : then this number determines an upper bound for the
dimension of the cohomology space of $S_L$ in the anomaly sector.

As a first result, this proves that if the manifold $\cal M$ has a complex
dimension smaller than 3, there is no anomaly candidate. Another special case
is the compact K\"ahler homogeneous one ( N=2 supersymmetric extension of our
previous work on the bosonic case \cite{7}) : in such a case the Ricci tensor
is positive definite \cite{77} which forbids (\cite{16},\cite{13}b)) the
existence of such analytic tensor $t^{[\al\be\ga]}(\phi^{\de})$. As a
consequence, the cohomology of $S_L^0$ - and then of $S_L$ - vanishes in the
anomaly sector (for details, see ref.\cite{13}b)).
$$ $$
We are now in a position to discuss the cohomology of the complete $S_L \equiv
S_L^0 + S_L^r$ operator.

\subsection{$S_L$ cohomology}
Thanks to the simplicity of $\De_{[-1]}$ ,the cohomology of the complete $S_L$
operator in the Faddeev-Popov charge -1 sector is easily obtained : the vector
$V^i[\phi]$ should satisfy :
$$\bullet \ \  \int d^2x d^2\th \frac{\de S^{inv.}}{\de
\phi^i(x,\th)}V^i[\phi(x,\th)] = 0 \ \Leftrightarrow V^i[\phi]\ {\rm is\  a\
Killing\ vector\ for\ the\ metric\ }\ g_{ij}[\phi]\ .$$
$$ \bullet J^i_j[\phi]\nabla_iV^k = \nabla_jV^iJ^k_i[\phi] \ \Leftrightarrow
V^i[\phi]\ {\rm  is\ a\ contravariant\ vector,\ analytic\ with\ respect\ to\ }\
 J^i_j[\phi]\ .$$
$$ $$
Consider now the Faddeev-Popov neutral charge sector. We shall prove that,
despite the non-vanishing  $S_L^0$ cohomology in a Faddev-Popov positively
charged sector (\ref{a92}), the cohomology of $S_L$ is a subspace of the one of
$S_L^0$, {\it i.e.} that one can always construct the cocycles for $S_L$
starting from those of $S_L^0$. Indeed, notice that $S_L^r \De_{[0]}$ contains
at most one source $\eta_i$ ; then it cannot intercept $\De^{an.}_{[+1]}$, the
cohomology of $S_L^0$ in the anomaly sector. As a consequence
(\cite{7},\cite{13}a)), there will be no obstruction in the construction of the
cocycles of $S_L$ starting from those of $S_L^0$.\footnote{\ Let us sketch the
proof. Under the filtration,$$ S_L \stackrel{(\ref{a5})}{=} S_L^0 +S_L^1 +S_L^2
+..\ \ \ \De_{[0]} \stackrel{(\ref{a81})}{=} \De_{[0]}^1 + \De_{[0]}^2
+\De_{[0]}^3 + ...$$
The cocycle condition $S_L\De_{[0]} = 0$ gives, using (\ref{a811}) :
$$S_L^0\De_{[0]}^1 = 0 \ \Rightarrow \ \De_{[0]}^1 = S_L^0\De_{[-1]}^1\ .$$
At the next step,
$$S_L^0\De_{[0]}^2 + S_L^1\De_{[0]}^1 = 0 \ \Rightarrow \ S_L^0[\De_{[0]}^2 -
S_L^1\De_{[-1]}^1] = 0 \ \Rightarrow \ \De_{[0]}^2 = \De_{[0]}^{an.\;2} +
S_L^0\De_{[-1]}^2 + S_L^1\De_{[-1]}^1\ .$$
At the next step,
$$S_L^0\De_{[0]}^3 + S_L^1\De_{[0]}^2 + S_L^2\De_{[0]}^1 = 0 \ \Rightarrow \
S_L^0[\De_{[0]}^3 - S_L^2\De_{[-1]}^1 - S_L^1\De_{[-1]}^2] +
S_L^1\De_{[0]}^{an.\;2} = 0 $$
where we have used $ S_L^0 S_L^2 + S_L^1 S_L^1 + S_L^2 S_L^0 = 0 $. The last
equation implies, using (\ref{a91}) $$  S_L^0( S_L^1\De_{[0]}^{an.\;2}) = 0 \
\Rightarrow \  S_L^1\De_{[0]}^{an.\;2} = \De_{[+1]}^{an.\;3} + S_L^0
\tilde\De_{[0]}^{3}$$
but, thanks to the upper remark, the anomalous term $\De_{[+1]}^{an.\;3}$
cannot appear here, and finally we get
$$S_L^0[\De_{[0]}^3 + \tilde\De_{[0]}^3 - S_L^2\De_{[-1]}^1 -
S_L^1\De_{[-1]}^2] = 0 $$
so that, using (\ref{a811}) $$\De_{[0]}^3 = - \tilde\De_{[0]}^3 +
S_L^2\De_{[-1]}^1 + S_L^1\De_{[-1]}^2 + S_L^0\De_{[-1]}^3 + \De_{[0]}^{an.\;3}
.$$
Finally, up to that order,
$$\De_{[0]}|_3 = - \tilde\De_{[0]}|_3 + (S_L\De_{[-1]})|_3+ \De_{[0]}^{an.}|_3
.\ \ e.t.c.\ ...\ \ Q.E.D.$$ }

It may however happen that some of the so doing constructed cocycles for  $S_L$
become coboundaries : this occurs when there is some cohomology for  $S_L^0$ in
the Faddeev-Popov charge -1 sector (\cite{13}a),\cite{17}). We have previously
seen that this relies on the existence of Killing vectors for the metric
$g_{ij}[\phi]$; this is natural as such vectors signal extra isometries that
constrain the invariant action (\ref{a1}) or equivalently, signal the non
physically relevant character of some of the parameters of the classical action
that may be reabsorbed through a conveniently chosen field and source
reparametrisation \cite{7}. Up to this restriction, the cohomology in the
Faddeev-Popov neutral sector is then characterized by a symmetric tensor
$t_{ij}[\phi]$ such that $g'_{ij} = g_{ij} + \hbar t_{ij}$ is a metric,
hermitian with respect to the very \footnote{\ As previously mentionned
(subsection 3.1), a reparametrisation of the sources and fields has been used
to compensate for the change $U^i_j[\phi]$ in the original complex structure
$J^i_j$.} complex structure $J^i_j$ we started from, and such that $J^i_j$ is
covariantly constant with respect to the covariant derivative with connexion
$\Ga^k_{ij}[g'_{mn}]$. This is the necessary stability of the theory which
ensures that, at a given perturbative order where the Slavnov identity holds (
absence of anomaly up to this order), the U.V. divergences in the Green
functions may be compensated-for through the usual renormalisation algorithm
and normalisation conditions \cite{6}. Of course, the trivial cohomology $ S_L
\De_{[-1]}[V^i(\phi)]  $ corresponds to field and source reparametrisations
according to (\ref{a40},\ref{a41}).

$$ $$
	Let us finally study the Faddeev-Popov charge +1 sector.
As in this letter we restrict ourselves to compact K\"ahler Ricci-flat
manifolds, if the Hodge number $h^{(3,0)} = h^{(0,3)}$ does not vanish, we have
a true anomaly candidate.
Starting from the $S_L^0$ cohomology (\ref{a92}) :
$$\De_{[+1]}^{an.} = \int d^2x d^2\th t^{[ijk]}[\phi](d^+)^2(d^-)^2
\eta_i\eta_j\eta_k \ \ ,$$

\noindent we were able to construct the $S_L$ cohomology in the same
Faddeev-Popov sector (\cite{13}b)) :
\beqa\label{a12}
\De_{[+1]}^{an.} &=& \int d^2x d^2\th [t^{[ijk]}[\phi]
\{(d^+)^2(d^-)^2\eta_i\eta_j\eta_k \nnb\\
&-& {3\over 2}d^+d^-(\eta_i\eta_j J_{kn}(d^+D_+\phi^n - d^-D_-\phi^n) + 2\eta_i
J_{jn} J_{km}D_+\phi^nD_-\phi^m )\nnb\\
&+& {3\over 4}  J_{in} J_{jm} J_{kl}(d^+D_+\phi^n D_+\phi^m D_-\phi^l -
d^-D_-\phi^n D_-\phi^m D_+\phi^l) \} \nnb\\
&+& \tilde {t}_{[nm]l}(d^+D_+\phi^n D_+\phi^m D_-\phi^l - d^-D_-\phi^n
D_-\phi^m D_+\phi^l)]
\eeqa

where $\tilde{t}_{[ij]\,k}[\phi]$ is related to $t^{[ijk]}[\phi]$ through (in
complex coordinates) :
$$\tilde{t}_{[\al\be]\,\bar{\ga}},\ \ \tilde{t}_{[\bar{\al}\bar{\be}]\,\ga}
\neq 0,\ \ {\rm \ the\ other\ vanish }\ ;$$
$$\tilde{t}_{[\al\be]\,\bar{\ga}} = -{i\over 4}
q\partial_{\bar{\ga}}[g_{\al\bar{\al}}
g_{\be\bar{\be}}K,_{\bar{\de}}
t^{[\bar{\al}\bar{\be}\bar{\de}]}] \ \
{\rm where\ K\ is\ the\ Kahler\ potential}\ .$$

\noindent As a consequence, if at a given pertubative order this anomaly
appears with a non zero coefficient
$$S_L\Ga|_{p^{th} order} = a (\hbar)^p \De_{[+1]}^{an.}, \ \ a \neq 0 $$
the N = 2 supersymmetry is broken as $\De_{[+1]}^{an.}$ cannot be reabsorbed
(being a  cohomology element, it is not a $S_L \tilde{\De}_{[0]}$ ) and, {\it a
priori}, we are no longer able to analyse the structure of the U.V. divergences
at the next perturbative order, which is the death of the theory.

\section {Concluding remarks}

In this letter we have analysed the cohomology of the B.R.S. operator
associated to N = 2 supersymmetry in a N = 1 superfield formalism. We have
found an anomaly candidate for torsionless models built on compact K\"ahler
Ricci-flat target spaces with a non vanishing Hodge number  $h^{3,0} =
h^{0,3}$. Calabi-Yau manifolds (3 complex dimensional case) where $h^{3,0} = 1
$ \cite{2} are interesting examples \footnote{\ As $\det\|g\| = 1$, a
representative of $t^{[\al\be\ga]}$ is the constant antisymmetric tensor
$\epsilon^{[\al\be\ga]}$( with $\epsilon^{123} = +1$).}
due to their possible relevance for supertring theories. Of course, as no
explicit metric is at hand, one can hardly compute the anomaly coefficient.

This anomaly in \underline{global} extended supersymmetry is a surprise with
respect to common wisdom \cite{18} ( but see other unexpected cohomologies  in
supersymmetric theories, in the recent works of Brandt \cite{8} and Dixon
\cite{9}) and the fact that if we have chosen, from the very beginning, a
coordinate system adapted to the complex structure, the second supersymmetry
will be linear and there will be no need for sources $\eta_i$ . However, as
known from chiral symmetry, even a linearly realised transformation can lead to
anomalies ; moreover, here the linear susy transformations do not correspond to
an ordinary group but rather to a supergroup where, contrarily to ordinary
compact groups \footnote{\ In the appendix A of ref. \cite{7}, it is proven
that any linearly realised symmetry corresponding to a \underline{compact}
group of transformations can be implemented to all orders of perturbation
theory.}, no general theorems exists : then there is no obvious contradiction.
This emphasizes the special structure of the supersymmetry algebra.

Of course, our analysis casts some doubts on the validity of the previous
claims on U.V. properties of N=2 supersymmetric non linear $\si$ models :
there, the possible occurence at 4-loops order of (infinite) counterterms
non-vanishing on-shell, even for K\"ahler Ricci-flat manifolds, did not ``
disturb" the complex structure ; on the other hand, we have found a possible ``
instability" of the second supersymmetry, which confirms that there are some
difficulties in the regularisation of supersymmetry by dimensional reduction
assumed as well in explicit perturbative calculations \cite{4} than in
finiteness ``proofs" \cite{3} or higher order counterterms analysis \cite{5}.
We would like to emphasize the difference between Faddeev-Popov 0 charge
cohomology which describes the stability of the classical action against
radiative corrections ( the usual ``infinite" counterterms) and which offers no
surprise, and the anomaly sector which describes the ``stability" of the
symmetry ( the finite renormalisations which are needed, in presence of a
regularisation that does not respect the symmetries of the theory, to restore
the Ward identities) : of course, when at a given perturbative order the
Slavnov (or Ward) identities are spoiled, at the next order, the analysis of
the structure of the divergences is no longer under control.
In particular, the Calabi-Yau uniqueness theorem for the metric \cite{19}
supposes that one stays in the same cohomology class for the K\"ahler form, a
fact which is not certain in the absence of a regularisation that respects the
N=2 supersymmetry (the possible anomaly we found expresses the impossibility to
find a regularisation that respects all the symmetries of these theories).

Of course, if one has added from the very beginning extra geometrical (or
physical !) constraints that would fix the classical action, we bet that our
anomaly candidate would disappear: as previously mentioned, this is the case
when the manifold is a compact \underline{homogeneous} K\"ahler space ;
moreover we have also been able to prove that, if one enforces N=4
supersymmetry (HyperK\"ahler manifolds), there is no possible tensor
$t^{[ijk]}[\phi]$ and then no anomaly (\cite{13}b)). Our final conjecture is
that the requirement of conformal invariance of the theory would be sufficient
to rule out a possible anomaly. We hope to be able to report on that subject in
a near future.

\bibliographystyle{plain}

\begin {thebibliography}{29}

\bibitem{1} P. Candelas, G. Horowitz, A. Strominguer and E. Witten,  {\sl Nucl.
Phys. } {\bf B258} (1985) 46.
\bibitem{1a} M. Green, J. Schwartz and E. Witten, ``Superstring Theory",
Cambridge University Press, and references therein.
\bibitem{2} G. Horowitz, `` What is a Calabi-Yau space ", in {\sl Worshop on
Unified String Theory, 1985} eds. M. Green and D. Gross (World Scientific,
Singapore), p. 635.
\bibitem{3} L. Alvarez-Gaum\'e, S. Coleman and P. Ginsparg, {\sl Comm. Math.
Phys.} {\bf 103} (1986) 423;\newline  C. M. Hull, {\sl Nucl. Phys.} {\bf B260}
(1985) 182.
\bibitem{4} M. T. Grisaru, A. E. M. van de Ven and D. Zanon, {\sl Phys. Lett. }
{\bf 173B} (1986) 423 ;\newline  M. T. Grisaru, D. I. Kazakov and D. Zanon,
{\sl Nucl. Phys.} {\bf B287} (1987) 189.
\bibitem{5} M. D. Freeman, C. N. Pope, M. F. Sohnius and K. S. Stelle, {\sl
Phys. Lett.} {\bf 178B} (1986) 199.
\bibitem{6} G. Bonneau, {\sl Int. Journal of Mod. Phys. }{\bf A5} (1990) 3831.
\bibitem{62}  W. Siegel, {\sl Phys. Lett.} {\bf 84B} (1979) 193 ; {\sl Phys.
Lett.} {\bf 94B} (1980) 37.
\bibitem{61} A. Galperin, E. Ivanov, S. Kalitzin, V. Ogievetsky and E.
Sokatchev, {\sl Class. Quantum Grav.} {\bf 1} (1984) 469.
\bibitem{63} A. Galperin, E. Ivanov, S. Kalitzin, V. Ogievetsky and E.
Sokatchev, {\sl Class. Quantum Grav.} {\bf 2} (1985) 601 ; 617 ; A. Galperin,
Nguyen Anh Ky and E. Sokatchev, {\sl Mod. Phys. Lett.} {\bf A2} (1987) 33.
\bibitem{7} C. Becchi, A. Blasi, G. Bonneau, R. Collina and F. Delduc, {\sl
Comm. Math. Phys.} {\bf 120} (1988) 121.
\bibitem{13} G. Bonneau, a) `` B.R.S. renormalisation of some on shell closed
algebras of symmetry transformations : the example of non-linear $\si$ models :
1) the N=1 case", LPTHE-Paris preprint PAR/LPTHE/94/10 ;
 \newline b) `` B.R.S. renormalisation of some on shell closed algebras of
symmetry transformations : 2) the N=2 and N=4 supersymmetric non-linear $\si$
models ", LPTHE-Paris preprint PAR/LPTHE/94/11.
\bibitem{8} F. Brandt, {\sl Nucl. Phys.} {\bf B392} (1993) 928.
\bibitem{9} J. A. Dixon, `` The search for supersymmetry anomalies - Does
supersymmetry break itself?", talk given at the 1993 HARC conference, preprint
CTP-TAMU-45/93 ({\sl and references therein}).
\bibitem{10} O. Piguet and A. Rouet, {\sl Nucl. Phys.} {\bf B99} (1975) 458.
\bibitem{100} D. Friedan,{\sl Phys. Rev. Lett. } {\bf 45} (1980) 1057 ,{\sl
Ann. Phys. (N.Y.)} {\bf 163} (1985) 318.
\bibitem{11} B. Zumino, {\sl Phys. Lett.} {\bf 87B} (1979) 203 ;\newline L.
Alvarez-Gaum\'e and D. Z. Freedman, {\sl Comm. Math. Phys.} {\bf 80} (1981)
443.
\bibitem{14} E. C. Zeeman, {\sl Ann. Math.} {\bf66} (1957) 557 ; J. Dixon,
``Cohomology and renormalisation of gauge fields", Imperial College preprints
(1977-1978)
\bibitem{151} G. Bonneau and G. Valent, ``Local heterotic geometry in
holomorphic coordinates", preprint PAR/LPTHE/93-56, to be published in {\sl
Class. Quantum Grav.}, in particular equation (16).
\bibitem{77} M. Bordemann, M. Forger and H. R\"omer, {\sl Comm. Math. Phys.}
{\bf 102} (1986) 605.\newline A. L. Besse, ``Einstein Manifolds", {\sl
Springer-Verlag. Berlin, Heidelberg, New-York} (1987).
\bibitem{16} K. Yano, Differential geometry on complex and almost complex
spaces, Pergamon Press (1965), and references therein.
\bibitem{17} C. Becchi and O. Piguet, {\sl Nucl. Phys.} {\bf B347} (1990) 596.
\bibitem{18} O. Piguet, M. Schweda and K. Sibold, {\sl Nucl. Phys.} {\bf B174}
(1980) 183 ; ``Anomalies of supersymmetry : new comments on an old subject",
preprint UGVA-DPT 1992/5-763.
\bibitem{19} S. T. Yau, {\sl Proc. Natl. Acad. Sci.} {\bf 74} (1977) 1798.





\end {thebibliography}

\end{document}